\documentstyle[preprint,aps]{revtex}
\draft
\begin{document}
\title{AC Josephson Effect in a Single Quantum Channel }
\author{D. Averin and A. Bardas}
\address{Department of Physics, SUNY Stony Brook, NY 11794}
\maketitle

\begin{abstract}
We have calculated all the components of the current in a short
one-dimensional channel between two superconductors for arbitrary
voltages and transparencies $D$ of the channel. We demonstrate that
in the ballistic limit ($D\simeq 1$), the crossover between the
quasistationary evolution of the Josephson phase difference $\varphi$
at small voltages and transport by multiple Andreev reflections at
larger voltages can be described as the Landau-Zener transition
induced by finite reflection in the channel. For perfect transmission
and vanishing energy relaxation rate the stationary current-phase
relation is never recovered, and $I(\varphi)=I_c \mid \sin \varphi /2
\mid \mbox{sign} V$ for arbitrary small voltages.

\end{abstract}

\narrowtext

It has been known for more than 20 years that electron transport
in short superconducting weak links with arbitrary transparency can
be described in terms of multiple Andreev reflections (MAR).\cite{b1}
Despite this, quantitative understanding of the ac Josephson effect
in these structures is still not complete. Various approaches to
quantitative calculations of the current at finite
voltages \cite{b2,b3,b4,b5} were mainly focused on the dc current
which exhibits the so-called subharmonic structure, i.e. current
singularities at voltages $V= 2\Delta /en, \; n=1,2,...,$ where
$\Delta$ is the superconducting energy gap -- see, e.g., \cite{b6,b7}
and references therein. However, dc current carries only indirect
information about weak link dynamics, whereas calculations of
the ac currents \cite{b8,b9} have been limited to large voltages, the
limitation being caused by the fact that at small voltages it is
necessary to take into account increasingly large number of Andreev
reflections.

The aim of our work was to study a model of a short
constriction between two superconductors which permits the
quantitative description of the current dynamics for
arbitrary voltages and transparencies of the constriction.
In this model, we found a new regime in the constriction
dynamics, which occurs at small voltages and connects quasistatic
variations of Josephson phase difference at $V\rightarrow 0$ with
MAR at larger voltages.

We consider a single-mode channel of electron gas with transparency
$D$ between two superconductors (the calculations can be generalized
in a straightforward way to several separable modes). The length $d$
of the channel is assumed to be much smaller than the coherence
length $\xi$ as well as elastic and inelastic scattering lengths in
the superconductors. This allows us to neglect scattering in the
vicinity of the channel (besides that described by the reflection
probability $R=1-D$) and makes it convenient to describe
electron motion in the constriction with the time-dependent
Bogolyubov-de Gennes
(BdG) equations. Assuming that the Fermi energy in the constriction
is much larger than the energy gap $\Delta$ we simplify these
equations further by adopting the quasiclassical approximation.
Condition $d\ll \xi$ makes the superconducting properties of the
constriction itself irrelevant (even if there is a finite $\Delta$
in the constriction we can neglect it in the BdG equations on the
small space scale given by $d$) \cite{b10}. It is easier to visualize
electron motion in the channel assuming that the constriction
is normal ($\Delta =0$), so that the transport through the
resulting SNS structure can be described directly in terms of
the Andreev reflection at the two NS interfaces. We adhere to this
framework in what follows.

The final simplification is that impedance of the single- (or few-)
mode channel is on the
order of $h/e^2$ and is much larger than characteristic impedance
of a typical external circuit. This eliminates the necessity
(essential for a realistic description of the Josephson junctions
with low resistance) of determining the dynamics of the Josephson
phase difference and voltage across the channel self-consistently.
We assume that the voltage is constant in time.

The model we obtain is directly applicable to the atomic-size
Josephson junctions\cite{b11} which exhibit ballistic
quantization of the
stationary critical current\cite{b12}. Another context of current
interest in which the model is relevant is high-critical-current
Josephson junctions \cite{b6,b13} which are believed to be
adequately represented as an ensemble of atomic-size
microconstrictions each of which carries a few conducting
electron modes.

Under the assumptions outlined above, the BdG equations for
transport in the constriction can be solved in terms of the two
scattering processes for electrons and holes, along the same lines
as in the stationary case\cite{b14}. One process is Andreev
reflection at the NS interfaces characterized by the reflection
amplitude $a$ as a function of the quasiparticle energy $\epsilon$:
\begin{equation}
a(\epsilon)=\frac{1}{\Delta} \left\{ \begin{array}{ll}
\epsilon- \mbox{sign}(\epsilon) (\epsilon^2- \Delta^2)^{1/2}\, ,
\;\;\; & \mid \epsilon \mid > \Delta\, , \\
\epsilon-i(\Delta^2 -\epsilon^2)^{1/2}\, , \;\;\; & \mid \epsilon
\mid < \Delta\, . \end{array} \right.
\label{1} \end{equation}
Another process is electron scattering in the constriction
characterized by a scattering matrix:
\begin{equation}
S_{el}= \left( \begin{array}{cc} r & t \\ t & -r^*t/t^*
\end{array} \right) \, ,
\label{2}  \end{equation}
where $\mid t \mid^2 =D$ and $\mid r \mid^2 =R$. The scattering
matrix for holes is the time reverse of $S_{el}$, $S_h=S_{el}^*$.

The last ingredient of the scattering scheme is the fact that the
energy of an electron is increased by $eV$ each time it passes
through the channel from left to right, while the hole increases
its energy passing through the constriction in the opposite
direction. Because of this the electron and
hole wave functions are sums of the components with
different energies shifted by $2eV$. For instance, the wave functions
in region I (Fig.\ 1) generated by the quasiparticle incident
from the left superconductor onto the channel can be written as:
\[ \psi_{el} =\sum_{n}[(a_{2n}A_n+J\delta_{n0})e^{ikx}+B_ne^{-ikx} ]
 e^{-i(\epsilon +2neV)t/\hbar }\, , \]
\begin{equation}
\psi_{h} =\sum_{n}[A_ne^{ikx}+a_{2n}B_ne^{-ikx} ]
e^{-i(\epsilon +2neV)t/\hbar } \, ,
\label{3} \end{equation}
where $k$ and $\epsilon$ are momentum (equal to the Fermi momentum)
and energy of the incident quasiparticle, and $a_m\equiv
a(\epsilon+meV)$. In eq.\ (\ref{3}) we took into account the fact
that the amplitudes of electron and hole waves are related by the
Andreev reflection, and the quasiparticle incident from
the superconductor produces an electron in the normal region with
amplitude $J(\epsilon)=(1-\mid \! a(\epsilon) \! \mid ^2)^{1/2}$.
The wave function in region II has a similar form with two
modifications: it does not have the source term $J$ and is shifted
in energy by $eV$.

The wave amplitudes in regions I and II are related by the
scattering matrix (\ref{2}):
\begin{equation}
\left( \begin{array}{c} B_n \\ C_n \end{array} \right) =
S_{el} \left( \begin{array}{c} a_{2n}A_n+J\delta_{n0} \\
a_{2n+1}D_n \end{array} \right) , \;\;\;\;\;\;
\left( \begin{array}{c} A_n \\ D_{n-1} \end{array} \right) =
S_{h} \left( \begin{array}{c} a_{2n}B_n \\
a_{2n-1}C_{n-1} \end{array} \right) \, .
\label{4}  \end{equation}
Eliminating the wave amplitudes $C_n,\, D_n$ in region II from
eq.\ (\ref{4}) we obtain the recurrence relation for the amplitudes
$A_n, \, B_n$:
\[ D\frac{a_{2n+2}a_{2n+1}}{1-a_{2n+1}^{2}} B_{n+1} -
[D(\frac{a_{2n+1}^2}{1-a_{2n+1}^{2}} + \frac{a_{2n}^2}{1-a_{2n-1}
^{2}})+1-a_{2n}^{2}] B_n +D\frac{a_{2n}a_{2n-1}}{1-a_{2n-1}^{2}}
B_{n-1} = - R^{1/2}J\delta_{n0} \, , \]
\begin{equation}
A_{n+1}-a_{2n+1}a_{2n}A_n =R^{1/2}(B_{n+1}a_{2n+2}-B_na_{2n+1})
+Ja_1\delta_{n0} \, .
\label{5} \end{equation}

These recurrence relations can be solved with the method developed
in \cite{b15,b5}. The amplitudes $A_n,\, B_n$ of
the wave functions (\ref{3}) obtained in this way determine all
Fourier components of the current $I(t)$ in the channel:
\[ I(t)=\sum_k I_k e^{i2keVt/\hbar} \, . \]
Collecting contributions from the quasiparticles incident on the
channel from the two superconductors, and making use of the
fact that $A(-\epsilon,-V)=-A^*(\epsilon,V)$ and
$B(-\epsilon,-V)=B^*(\epsilon,V)$ (as follows from the recurrence
relations (\ref{5}) and the form of Andreev reflection amplitude
(\ref{1})) we finally arrive at:
\begin{eqnarray}
I_k =\frac{e}{\pi \hbar} \left[ eV \delta_{k0} - \int d\epsilon
\tanh \{ \frac{\epsilon}{2T} \} (J(\epsilon )(a_{2k}A_k^* +a_{-2k}
A_{-k}) \right. \nonumber \\ \left. + \sum_n (1+a_{2n}a_{2(n+k)}^*)
(A_n A_{n+k}^* -B_n B_{n+k}^*)) \right]  \, .
\label{6} \end{eqnarray}

Some results of the numerical calculations of the current from the
recurrence relations (\ref{5}) and equations (\ref{6}) are shown
in Fig.\ 2. One can see that ac components of the current exhibit
the subgap singularities at $V=2\Delta/n$ similar to those in the
dc current. It is clear from Fig.\ 2 (and straightforward to
show analytically) that in the limit $D\rightarrow 0$ eqs.\
(\ref{5}), (\ref{6}) reproduce the tunnel Hamiltonian expressions
for the current. In the case of zero temperature (shown in Fig.\ 2)
the only component of the current that is non-vanishing at finite
reflection probability and small voltages is the stationary
Josephson current. In particular, it can be checked that the limiting
($V=0$) values of the sine component (Fig.\ 2b) coincide with
the first Fourier harmonics of the stationary Josephson current
$I(\varphi)= (eD\Delta /2\hbar) \sin \varphi /(1-D \sin^2 (\varphi/2)
)^{1/2} $.

Figure 2c shows that for large transmission probabilities $D$
cosine component of the current is negative in the wide range of
voltages below the gap voltage $2\Delta/e$. (Calculations
based on eqs.\ (\ref{5}),(\ref{6}) at finite temperatures show
that this feature is preserved at temperatures up to about
0.3$\Delta$.) This fact provides a possible resolution of the
long-standing problem of sign of the cosine component of the ac
Josephson current in tunnel junctions (see, e.g., \cite{b15*}),
if one assumes that due to non-uniformity of tunnel barriers
realistic tunnel junctions always contain regions with high
transparency.

The feature of the curves in Fig.\ 2 which to our knowledge
has never before been discussed is the rapid variation of all current
components at small voltages and small reflection probabilities.
In order to understand this new feature we consider first the
case of perfect transmission, $D=1$. In this case the recurrence
relations can be solved explicitly,
\begin{equation}
B_n=0, \;\;\; A_n=J \prod_{j=1}^{2n-1} a_{j} \, .
\label{11} \end{equation}
This solution shows that in the limit $V\rightarrow 0$ when
electrons and holes increase their energy in Andreev
reflections cycles very slowly, the amplitudes $A_n$ decrease
very rapidly as functions of energy outside the energy gap,
since $\mid a(\epsilon) \mid<1$ at these energies. At the
same time, the number of Andreev reflections
inside the gap increases. This means that, in this regime, the
dominant contribution to the current comes from the gap interval
$\mid \! \epsilon \! \mid < \Delta$, where $A_n$ are not decaying
($a(\epsilon) =\exp\{ -i \arccos(\epsilon/\Delta) \}, \,  \mid
a \mid=1$) and the number of Andreev reflection cycles diverges as
$\Delta/eV$. This divergence is regularized by the fact that the
particles getting into the energy gap originate only from a
small energy range near the gap edges. With this understanding
eq.\ (\ref{6}) for the $k$th Fourier component of the current can
be simplified for $V\ll \Delta/e$ as follows:
\begin{equation}
I_k = \frac{e\Delta}{\pi \hbar} \tanh (\frac{\Delta}{2T})
\int_{-1}^{1} dz \exp \{ 2ik\arccos z\} .
\label{12} \end{equation}
Equation (\ref{12}) means that the current-phase relation at finite
voltages is:
\begin{equation}
I(\varphi) =I_c \mid \sin \varphi/2 \mid \mbox{sign} V\, ,
\;\;\;\;\; I_c=\frac{e\Delta}{\hbar} \tanh (\frac{\Delta}{2T}) \, ,
\;\;\; \varphi = 2eVt/\hbar \, ,
\label{13} \end{equation}
in particular, the dc current is $2I_c/\pi$, the sine part of the
first Fourier component of the ac current is vanishing, while the
cosine part is $-2I_c/3\pi $.

Expression (\ref{13}) has a natural interpretation in terms of the
quasistationary discrete states inside the gap that are responsible
for the stationary Josephson current \cite{b16,b12}. Two such
states with energies $E_{\pm} =\Delta \cos \varphi/2$ (Fig.\ 3)
carry, respectively, forward and backward current, and in the
stationary case ($\varphi=\mbox{const}$) are occupied according to
equilibrium Boltzman distribution. At finite voltage the energy of
these states is changing in time due to evolution of $\varphi$. For
vanishing inelastic scattering rate the density of states in the
superconductors is also vanishing within the gap, so that the
occupation probabilities of the two current-carrying states
remain constant as long as they are moving inside the energy gap,
$E_{\pm}<\Delta$. The only point at which the occupation
probabilities can change is at the gap edges $\epsilon= \pm
\Delta$, (i.e. at $\varphi =2\pi n$). These considerations
immediately give eq.\ (\ref{13}).

Solution (\ref{11}) of the scattering problem at $D=1$ gives
some analytical results at larger voltages also. In
particular it can be shown from eqs. (\ref{6}),(\ref{11}) that
at $T=0$ the sine part of the first Fourier component of the
ac Josephson current
vanishes identically at all voltages, while cosine part
approaches the asymptote $-I_c (\Delta \ln 2/2eV) $ at
$eV\gg \Delta$. Both of these results agree with the
numerical results shown in Fig.\ 2.

The reasoning that lead to eq.\ (\ref{13}) can be easily
generalized to the small but finite reflection
probability $R$ of the constriction. Finite reflection creates
a finite matrix element $r$ of the transition between two
current-carrying states in the energy gap which occur near the
point $\varphi= \pi$, where the energies $E_{\pm}$ of these
states coincide -see Fig.\ 3. The problem of this transition
is then a standard level-crossing problem and the probability
$p$ that the system will continue to occupy the same level
after crossing the point $\varphi =\pi$ is given by the
Landau-Zener expression. In our notations this expression is:
\begin{equation}
p=\exp \{-\frac{\pi R \Delta}{eV} \} \, .
\label{14} \end{equation}

The finite transition probability $p$ modifies the current-phase
relation (\ref{13}) as follows:
\begin{equation}
I(\varphi) =I_c \mbox{sign} V \left\{ \begin{array}{ll}
\sin \varphi/2 \, , \;\;\;\; & 0< \varphi <\pi \, , \\
(2p-1)\sin \varphi/2 \, , \;\;\;\; & \pi < \varphi <2\pi \, .
\end{array} \right.
\label{15} \end{equation}

In the relevant range of parameters (small R and V) eq.\
(\ref{15}) reproduces the result of numerical solution of the
recurrence relations (\ref{5}). Indeed, at very small $V$
($eV<R\Delta$), $p\rightarrow 0$ and the system in its evolution
follows a current-phase relation which at $T=0$ coincides with
the stationary relation $I=I_c \sin \varphi/2 \, \mbox{sign}(
\cos \varphi/2 )$. As a result, the dc current and the cosine
component of the first harmonic of the ac current are vanishing,
while sine component is equal to its stationary value, in agreement
with Fig.\ 2. At larger voltages ($eV>R\Delta$), $p\rightarrow 1$
and the current-phase relation approaches the one for $D=1$,
$I(\varphi) =I_c \mbox{sign} V \mid \sin \varphi/2 \mid$. For this
$I(\varphi)$ the sine component is zero while cosine component and
dc current are non-vanishing. All this means that the reason for
the rapid variation of all current components with voltage at
$D\simeq 1$ and small voltages (see Fig.\ 2) is that the
probability (\ref{14}) of Landau-Zener transition between the two
current-carrying states changes rapidly on a small voltage scale
given by $R\Delta$.

Before concluding, we would like to mention that our
calculations agree with most of the previous results on the ac
Josephson effect in short constrictions, in the parameter ranges
where previous results are available. In particular, our
numerical results (Fig.\ 2) agree with the numerical results of
Arnold \cite{b9} for large voltages. At $D=1$ eq.\ (\ref{13}) of
our work gives the dc current in agreement with that obtained by
Gunsenheimer and Zaikin \cite{b4}. There is, however, a
contradiction between calculations at large voltages based
on the solution (\ref{11}) and Zaitzev's results for ac components
of the current at $D=1$ and large voltages \cite{b8}. The reason
for this contradiction is not clear at present.

In conclusion, we have calculated the current in a short single-mode
electron channel between two superconductors for arbitrary
voltages and transparencies of the channel. To the best of our
knowledge this is the first time when full description of the
current dynamics in a weak link is developed. In the ballistic limit,
$D\simeq 1$ crossover from quasistationary Josephson current at
smaller voltages to multiple Andreev reflections at larger voltages
occurs at $V\simeq \pi R\Delta/e$ and can be described in terms of
Landau-Zener tunneling between the discrete current-carrying states
in the energy gap which are responsible for the stationary Josephson
current at $V=0$.

The authors gratefully acknowledge numerous useful discussions with
M. Kupriyanov and K. Likharev. This work was supported by DOD URI
through AFOSR Grant \# F49620-92-J0508.

\figure{Figure 1. Schematic energy diagram of a short one-dimensional
channel with arbitrary transparency between two
superconductors. I and II denote the portions of the channel
separated by the scattering region. \label{f1}}

\figure{ Figure 2. DC current, and sine and cosine parts of the first Fourier
component of the ac current in a short quantum channel between
the two superconductors at zero temperature. $G_T$ is the
normal-state conductance of the channel, $G_T = e^2D/\pi \hbar$.
All curves exhibit
subharmonic singularities at $V=2\Delta /n$ associated with multiple
Andreev reflections. Rapid variation of the curves with $D\simeq 1$
at small voltages is a manifestation of the Landau-Zener transitions
between the quasistationary current-carrying states in the energy
gap. For details see text. \label{f2}}

\figure{ Figure 3. The energy diagram of the two quasistationary
current-carrying states $E_{\pm}$ in the constriction. The arrows
show two possible routes of the system evolution due to Landau-Zener
transitions in the vicinity of the level-crossing point
$\varphi=\pi$. \label{f3} }

\end{document}